\documentclass[runningheads]{llncs}
\usepackage[T1]{fontenc}
\usepackage{hyperref}
\usepackage{amsmath}
\usepackage{amssymb}
\usepackage{booktabs}
\DeclareMathOperator*{\argmax}{argmax}
\usepackage{wrapfig}
\usepackage{indentfirst}
\usepackage{mathtools}
\usepackage[caption=false, font=footnotesize]{subfig}
\usepackage{array,longtable,multicol,multirow,siunitx,tabularx}
\usepackage{graphicx}
\begin{document}
\title{Towards Unsupervised Speaker Diarization System for 
Multilingual Telephone Calls \\Using Pre-trained Whisper Model and \\
Mixture of Sparse Autoencoders}
%
\titlerunning{Unsupervised Speaker Diarization for Multilingual Calls}
%
\author{ Phat Lam\inst{1} 
\and Lam Pham\inst{2} 
\and Truong Nguyen \inst{1}
\and Dat Ngo \inst{3}
\and Thinh Pham \inst{4}
\and Tin Nguyen \inst{1}
\and Loi Khanh Nguyen \inst{1}
\and Alexander~Schindler \inst{2}
}
%
\authorrunning{P. Lam et al.}
%
\institute{Ho Chi Minh University of Technology, Vietnam \and
Austrian Institute of Technology, Austria
\and
University of Essex, United Kingdom
\and 
Ho Chi Minh City University of Science, Vietnam}
\maketitle     
\begin{abstract}
Existing speaker diarization systems typically rely on large amounts of manually annotated data, which is labor-intensive and difficult to obtain, especially in real-world scenarios. Additionally, language-specific constraints in these systems significantly hinder their effectiveness and scalability in multilingual settings.
In this paper, we propose a cluster-based speaker diarization system designed for multilingual telephone call applications. Our proposed system supports multiple languages and eliminates the need for large-scale annotated data during training by utilizing the multilingual Whisper model to extract speaker embeddings. Additionally, we introduce a network architecture called Mixture of Sparse Autoencoders (Mix-SAE) for unsupervised speaker clustering. Experimental results on the evaluation dataset derived from two-speaker subsets of benchmark CALLHOME and CALLFRIEND telephonic speech corpora demonstrate the superior performance of the proposed Mix-SAE network to other autoencoder-based clustering methods. The overall performance of our proposed system also highlights the promising potential for developing unsupervised, multilingual speaker diarization systems within the context of limited annotated data. It also indicates the system's capability for integration into multi-task speech analysis applications based on general-purpose models such as those that combine speech-to-text, language detection, and speaker diarization.


\end{abstract}
\section{Introduction}
\begin{figure}[t]
    \centering
    \includegraphics[width = 0.9\textwidth, height = 0.43\textwidth ]{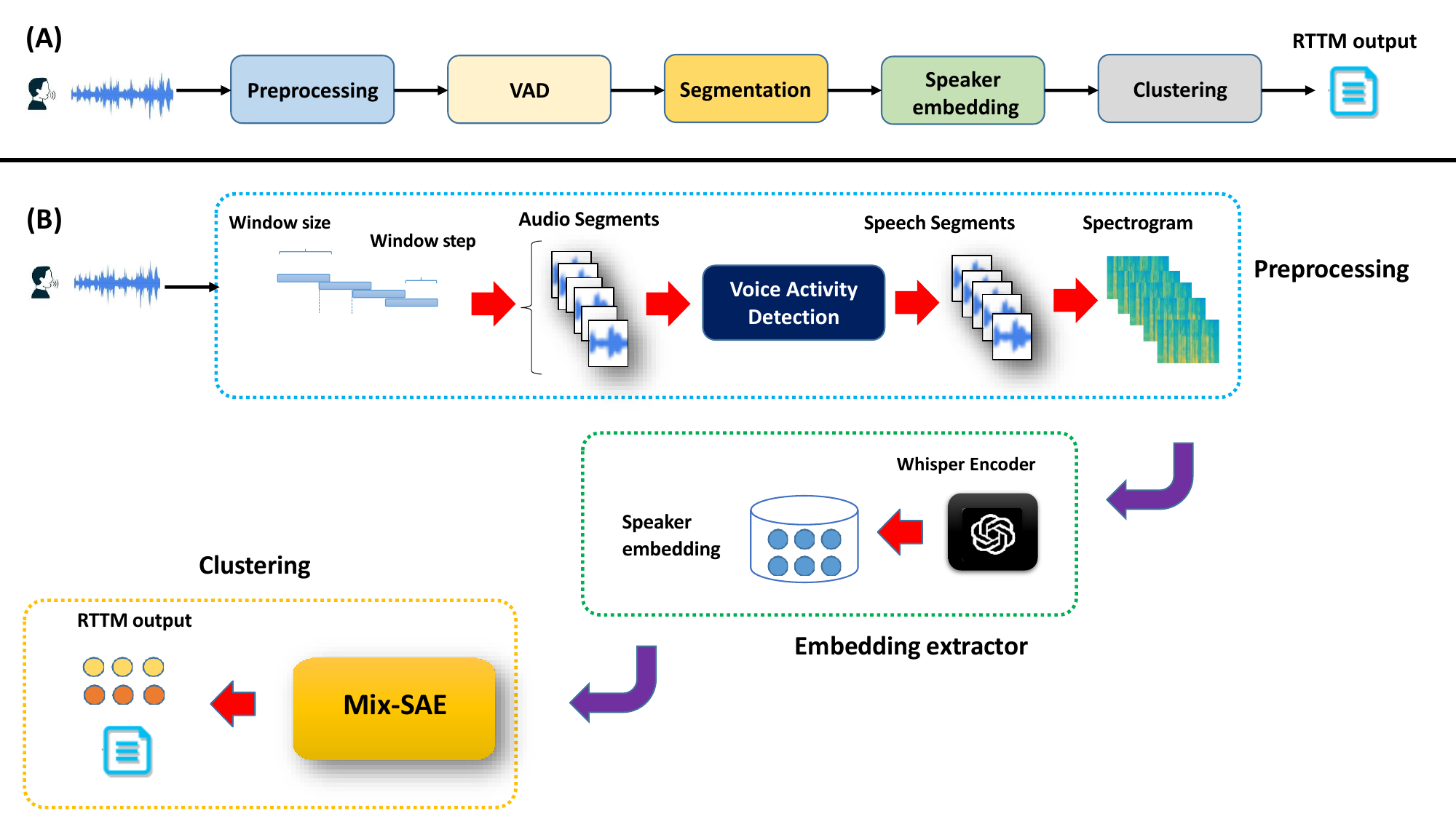}
    \caption{\textit{The high-level architectures of (A) Traditional cluster-based speaker diarization system and (B) Our proposed unsupervised speaker diarization system}} 
    \label{fig:overall}
    \vspace{-0.3cm}
\end{figure}

Sound-based applications have drawn significant attention from the research community and have become an integral part in the forefront of driving innovation. 
These applications involve advanced audio processing techniques to analyze and  interpret various types of sound data (e.g. acoustic scenes \cite{lightweight_asc},\cite{lampham_01}), sound events \cite{sed}, machinery sound \cite{tin_impact}, human speech  \cite{mofrad2018speech}), enabling the core functionality in many  intelligence systems. 
In human speech analysis, speaker diarization plays a crucial role by identifying and segmenting audio streams based on speaker identity, making it essential for various applications such as communication (e.g. customer support calls), security (e.g. voice tracking), healthcare (e.g. patient monitoring), smart home (e.g. personal assistants), etc. 
Typically, a cluster-based speaker diarization system consists of five modules. The traditional approach to such a system is illustrated at the top of Fig. \ref{fig:overall}. The preprocessing module first converts raw audio into a suitable format, followed by the voice activity detection (VAD) module extracting speech segments. These segments are then divided into fixed-length speaker segments. The speaker embedding extractor converts these segments into vectors representing speaker characteristics, and a clustering algorithm assigns speaker labels. Among these modules, speaker embedding and clustering are crucial components to enhance the performance of a cluster-based speaker diarization system \cite{spk_survey}.


Regarding the speaker embedding extractor, numerous approaches have been proposed for speaker embedding extraction, including metric-based models (GLR \cite{glr}, BIC \cite{bic}, etc),  probabilistic models (GMM-UBM \cite{gmm-ubm}, i-vectors \cite{ivectors}, etc), and neural network-based models (d-vectors \cite{dvectors}, x-vectors \cite{xvectors}, etc.). 
All these methods require a substantial amount of annotated data, especially for neural network-based approaches, to optimize  speaker feature extractors. 
However, training these extractors on one type of dataset could reduce the model's ability to generalize to diverse or unseen data, particularly from different domains.
In addition, datasets for speaker diarization mainly support one single language, due to the labor-intensive and time-consuming nature of collecting data and insufficient availability of data from diverse languages, limiting the effectiveness of speaker diarization systems in multilingual speech analysis applications. 

Concerning the clustering module, common methods such as Agglomerative Hierarchical Clustering (AHC) \cite{ahc_cluster}, k-Means \cite{kmeans_clustering}, Mean-shift \cite{mean_shift} have been proposed. However, these methods operate directly on the input vector space and rely heavily on distance-based metrics, without leveraging representation learning techniques to uncover deeper patterns. While some deep learning-based frameworks, such as DNN~\cite{dnn_based_cluster}, GAN~\cite{gan_cluster}, and Autoencoder~\cite{conv_autoencoder_cluster}, incorporate representation learning for speaker embeddings, they often require pre-extracted embeddings (e.g. x-vectors) that fit on certain datasets and are primarily evaluated on single-language datasets, typically English.


%
To address existing limitations, we aim to develop an unsupervised speaker diarization system that does not rely on large training datasets and supports multiple languages. For speaker embedding extraction, we use the multilingual Whisper model. This model is trained on diverse audio data for relevant tasks such as speech recognition, language identification, and translation. However, its applicability in speaker diarization task remains unexplored. Thus, leveraging Whisper's scalability and robustness, we explore its potential to produce high-quality speaker embeddings for diarization, assuming that as a general-purpose model, Whisper can learn representations that incorporate various aspects of large training data (e.g., phonetic content, acoustic features) that may be useful for diarization task, despite being primarily designed for speech recognition and speech-to-text transcription~\cite{xvectors}. For speaker clustering, we propose an unsupervised deep clustering network called Mixture of Sparse Autoencoders (Mix-SAE) to cluster the extracted embeddings.
Overall, our key contributions can be summarized as follows:
\begin{itemize}
    \item We explored the Whisper model's capability in the diarization task by using it as an alternative to conventional speaker embedding extractors, eliminating the need for annotated training data in developing diarization systems.
    \item Inspired by the work in \cite{damic}, we proposed the Mix-SAE network for speaker clustering, which enhances both speaker representation learning and clustering by using a mixture of sparse autoencoders with pseudo-label supervision.
    \item Through extensive experiments, we demonstrated that speaker diarization can be effectively integrated into
    Whisper-based systems, enabling comprehensive and multilingual speech analysis applications that combine speech-to-text, language identification, and speaker diarization.
    An example of a Whisper-based speech analysis application can be found at \footnote[1]{ \href{https://huggingface.co/spaces/AT-VN-Research-Group/SpeakerDiarization}{https://huggingface.co/spaces/AT-VN-Research-Group/SpeakerDiarization}}.
\end{itemize}

The remainder of this paper is organized as follows: The overall proposed speaker diarization system is described in Section 2. Next, Section 3 comprehensively describes our proposed deep clustering framework (Mix-SAE). Experimental settings and results are discussed in Section 4. The conclusion is represented in Section 5.
\section{The Overall Proposed System}
\label{pretrain}

Our proposed system pipeline is comprehensively described at the bottom of Fig. \ref{fig:overall}. 
Generally, the system comprises three main blocks: Front-end preprocessing, Speaker embedding extraction and Unsupervised clustering. The next subsections represent each block of the overall pipeline in detail.

\subsection{Front-end preprocessing}
Firstly, the input audio is divided into fixed-length segments of $W$ seconds and re-sampled to 16 kHz using Librosa toolbox~\cite{librosa}. To match the Whisper encoder's input requirements, zero-padding is applied to the segments. Next, a voice activity detection (VAD)~\cite{vad} is performed using an energy-based threshold to extract speech segments, which are then converted into spectrograms via Short-time Fourier Transform (STFT) with the setting of 400 filters, 10-ms window size, and a 160-sample hop size, respectively. These spectrograms are used as inputs to the Whisper encoder for speaker embedding extraction.
\begin{table}[t]
    \centering
    \caption{The Pre-trained Whisper Models}
    \label{tab:whispervar}
    \scalebox{0.8}{
\begin{tabular}{c|cc}
\hline
\textbf{Version} & \textbf{Parameters} & \textbf{Embedding Dimension} \\ \hline
Tiny                      & 39M                 & 384                        \\
Base                          & 74M                 & 512                        \\
Small                       & 244M                & 768                        \\
Medium                       & 769M                & 1024                       \\
Large                         & 1550M               & 1280 \\
\hline
\end{tabular}
    
    }
\end{table}
\subsection{Speaker embedding extraction using Whisper model}
In our work, we explore using the Whisper model as an alternative to conventional speaker embedding extractors, leveraging its scalability and diverse training data. We aim to utilize Whisper's robustness and generalization to capture various speaker characteristics across languages and domains. This approach allows us to obtain speaker embeddings directly from Whisper, bypassing the need for specific training datasets. For each speech segment, we generate the speaker embedding by feeding its spectrogram into the Whisper model. The final one-dimensional speaker embedding is derived by averaging the 2D tensor output from the last residual attention block of the Whisper encoder along the second axis, with its dimension varying by Whisper model versions, as shown in Table~\ref{tab:whispervar}.


\subsection{Unsupervised Clustering}
Given the speaker embeddings extracted from the Whisper model, the unsupervised clustering block groups together speech segments that are likely to be from the same speaker. In this work, we propose a new unsupervised deep clustering method called Mixture of Sparse Autoencoders (Mix-SAE). The proposed network uses a Mixture of Experts (MoE) architecture applied to Sparse Autoencoders (SAE), as detailed in Section~\ref{autoencoder}.
After clustering, we assign speakers to each segment and generate the diarization prediction by organizing the segments according to these assignments.
\section{Mixture of Sparse Autoencoder Deep Clustering Network (Mix-SAE)}
\label{autoencoder}
Our proposed Mix-SAE architecture, shown in Fig. \ref{fig:mae}, consists of two main parts: A set of k-sparse autoencoders, each representing a speaker cluster; and a gating projection that interprets the outputs produced by each autoencoder and assigns the input to a specific sparse autoencoder via its trainable weights.

\subsection{Individual Sparse Autoencoder (SAE)} \label{section_sae}
\begin{figure}[t]
    \centering
    \includegraphics[width = 0.7\textwidth, height = 0.35\textwidth]{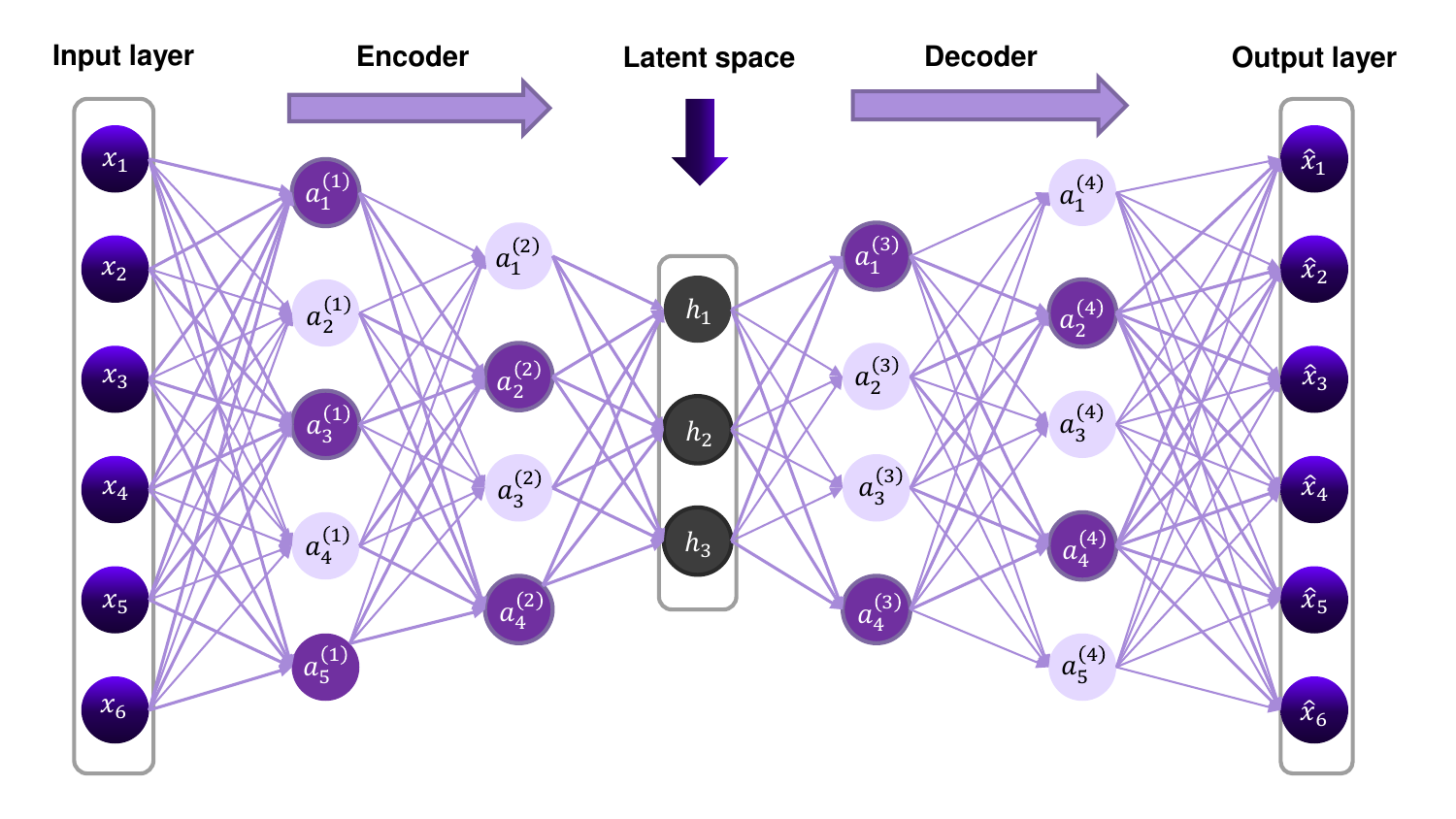}
    \caption{\textit{The Sparse Autoencoder Architecture (SAE)}}
    \label{fig:sae}
\end{figure}
Consider one sparse autoencoder $\mathcal{A}$, represented at Fig. \ref{fig:sae}. The sparse autoencoder $\mathcal{A}$ has $2L + 1$ layers, including one encoder ($\mathcal{E}$) with $L$ layers, one decoder ($\mathcal{D}$) with $L$ layer and one latent layer. We denote $a^{(l)}_{j}$ as the activation of hidden unit $j$ at the $l$-th hidden layer, $z_{j}^{(i)}$ is the input of $i$-th sample that leads to hidden unit $j$. Inspired from  \cite{andrewsparse}, we obtain the average activation of hidden unit $j$ at $l$-th layer over one batch of $N$ samples, which is written as:
\begin{equation}
    \hat{\rho}_{j}^{(l)} = \frac{1}{N} \sum^{N}_{i=1}\left[g\left(a^{(l)}_{j}(z^{(i)}_{j})\right)\right]
\end{equation}
where the mapping $g(.)$ uses the sigmoid function, which aims to scale the activation parameter to $[0;1]$ and avoid too large value of $\hat{\rho}_j^{(l)}$. The sparsity constraint ensures the average activation $\hat{\rho}_j^{(l)}$ is close to the sparsity parameter $\rho$, which is quite small. This helps the model learn meaningful features while avoiding copying or memorizing the input by enforcing a limited number of activation neurons in the hidden layer.
To achieve the approximation $\hat{\rho}_j \approx \rho$, we leverage Kullback–Leibler divergence penalty term \cite{andrewsparse}. 
The KL penalty term applied for the $l$-th hidden layer that has $n^{(l)}$ hidden units can be written as:
\begin{equation}
    \mathcal{L}^{(l)}_{\text{pen}} = \sum^{n^{(l)}}_{j=1} \text{KL}(\rho||\hat{\rho}_j^{(l)})=\sum^{n^{(l)}}_{j=1}\rho\log{\frac{\rho}{{\hat{\rho}}_j^{(l)}}} + (1-\rho)\log{\frac{1-\rho}{1-\hat{\rho}_j^{(l)}}}
\end{equation}
Then, the penalty term is calculated for all hidden layers of the autoencoder $\mathcal{A}$ (except the latent layer) by taking the sum of KL terms as:
\begin{equation}
    \mathcal{L}_{\text{pen}} =\sum^{2L}_{l=1}\sum^{n^{(l)}}_{j=1}\rho\log{\frac{\rho}{{\hat{\rho}}_j^{(l)}}} + (1-\rho)\log{\frac{1-\rho}{1-\hat{\rho}_j^{(l)}}}
    \label{kl}
\end{equation}

We also apply MSE loss for the pair of input data $\boldsymbol{x}$ and reconstruction data $\boldsymbol{\overline{x}}$ for one batch of $N$ samples as: 
\begin{equation}
\mathcal{L}_{\text{MSE}} = \frac{1}{2N}\sum^{N}_{i=1} ||\boldsymbol{x}_i-  \mathcal{D}\left(\mathcal{E}(\boldsymbol{x}_i)\right)||^2_2    
\label{mse}
\end{equation}

Given the KL penalty and MSE losses, we define the final objective function for the optimization of one individual sparse autoencoder $\mathcal{A}$:
\begin{equation}
    \mathcal{L}_{\text{SAE}} = \mathcal{L}_{\text{MSE}} + \beta \mathcal{L}_{\text{pen}}
    \label{total_loss}
\end{equation}
where $\beta$ is the parameter to control the effect of sparsity constraint on the objective function.


\subsection{k-Sparse Autoencoders}

\begin{figure}[t]
    \centering
    \includegraphics[width = 0.75\textwidth, height = 0.42\textwidth ]{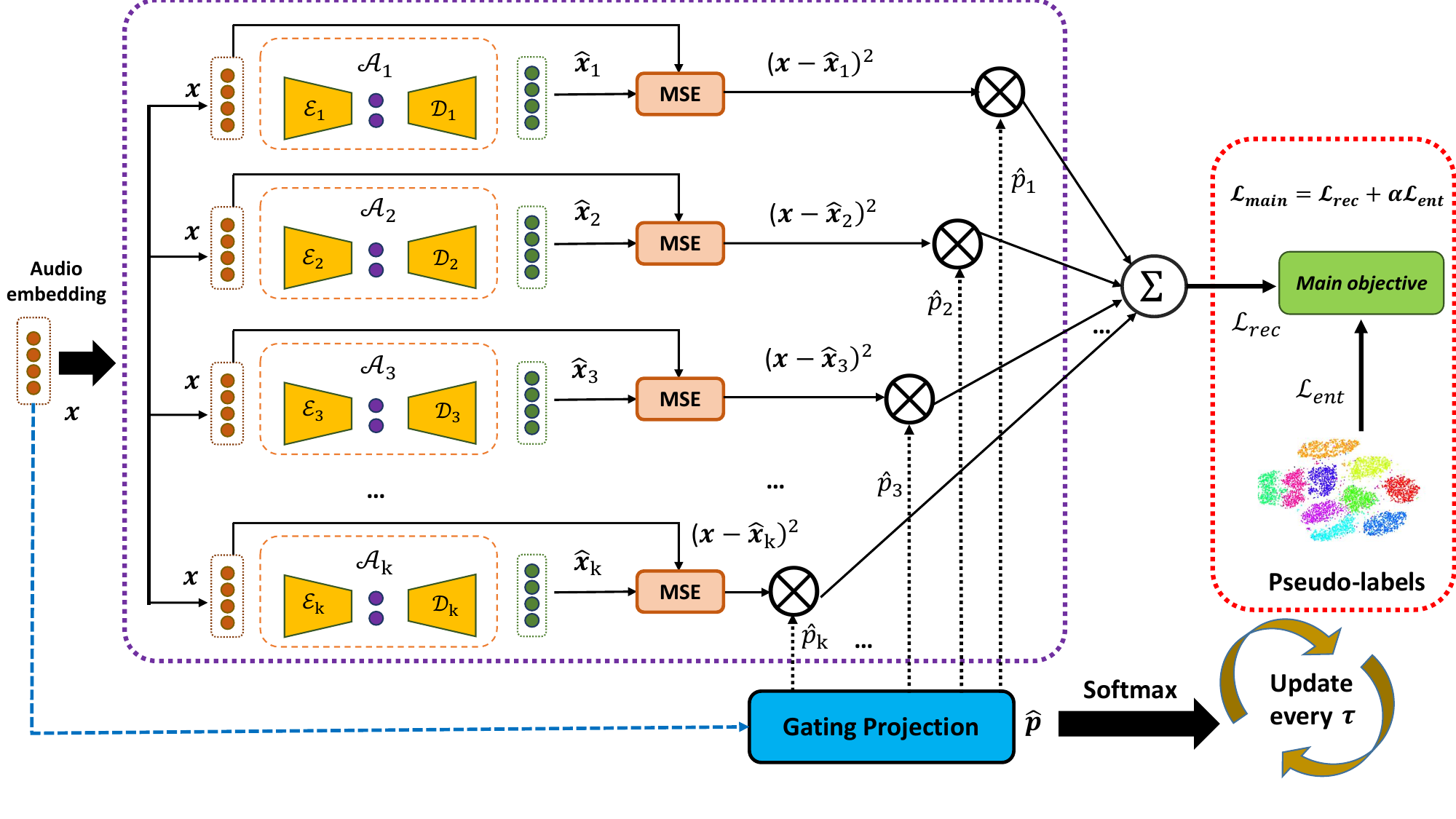}
    \caption{\textit{The overall architecture of Mix-SAE clustering network}} 
    \label{fig:mae}
\end{figure}

Given the problem of clustering a set of $M$ points $\{\boldsymbol{x}^{(i)}\}^{M}_{i=1} \in \mathbb{R}^m$ into $K$ clusters, the classical k-Means algorithm uses a centroid to represents each cluster in the embedding space, the centroids are mostly calculated by taking the average of all points belonging to that cluster. 
Inspired by \cite{damic} and \cite{kdae}, we use k-autoencoders to represent k clusters, with each autoencoder's latent space acting as a cluster centroid. In this paper, we use sparse autoencoders instead of standard ones, resulting in k-sparse autoencoders as shown in Fig.\ref{fig:mae}. This approach allows data points in the same cluster have their own autoencoder, making feature learning more efficient compared to using a single autoencoder for all data \cite{damic}. In our deep clustering network, all k-sparse autoencoders share the same settings and loss function $\mathcal{L}_{\text{SAE}}$ from Equation~\ref{total_loss}.

\subsection{Gating Projection} The role of the Gating Projection ($\mathcal{G}$) is to assign weights $\hat{\boldsymbol{p}} = [\hat{p}_1, \hat{p}_2, ..., \hat{p}_k]$ to the outputs of k-sparse autoencoders based on the input data.
Given the weights of $\hat{\boldsymbol{p}} = [\hat{p}_1, \hat{p}_2, ..., \hat{p}_k]$, the Gating Projection is also utilized to assign labels for clusters during the inference phase.
In this work, the Gating Projection leverages an MLP architecture with a single linear layer, followed by Leaky ReLU activation and the final softmax layer. Given the input data $\boldsymbol{x}$, the Gating Projection ($\mathcal{G}$) produces weights $\hat{\boldsymbol{p}} = [\hat{p}_1, \hat{p}_2, ..., \hat{p}_k]$ as:
\begin{equation}
    \hat{\boldsymbol{p}} = [\hat{p}_1, \hat{p}_2, ..., \hat{p}_k]=Softmax\left(\boldsymbol{W}\boldsymbol{x}+\boldsymbol{b}\right) \in \mathbb{R}^k 
    \label{gate}
\end{equation}
where $\boldsymbol{W} \in \mathbb{R}^{k\times m},\, \boldsymbol{b} \in \mathbb{R}^{k}$ are the trainable weights and bias of the linear layer in the gating projection. 

\subsection{Training strategy} 
The training strategy for our proposed Mix-SAE clustering network includes two steps: Pre-training and Main-training.

\begin{figure}[t]
    \centering
\includegraphics[width = 0.7\textwidth, height = 0.39\textwidth ]{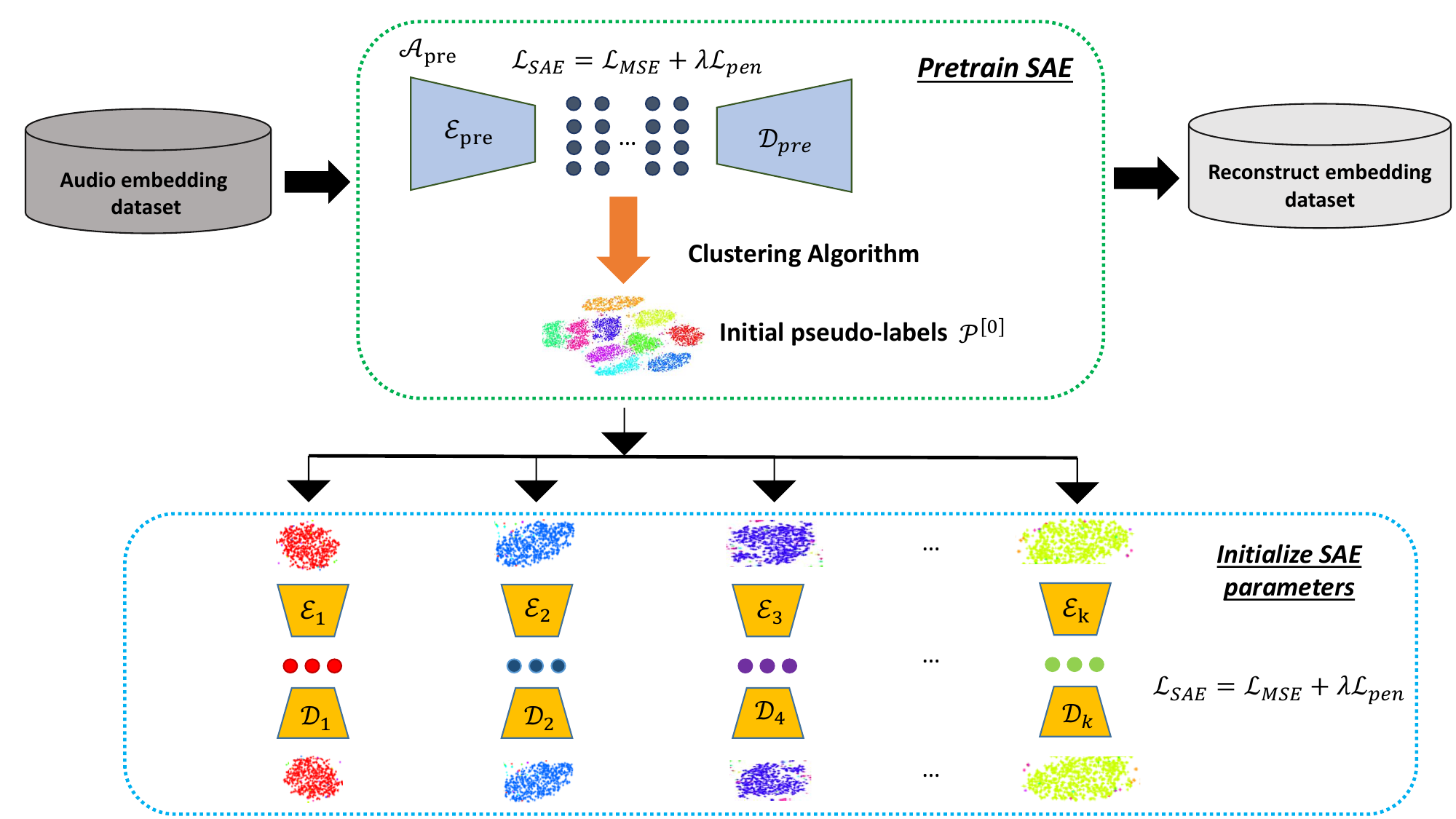}
    \caption{\textit{The Pre-training step of Mix-SAE clustering network}} 
    \label{fig:pretrain}
\end{figure}

In the Pre-training step as shown in Fig. \ref{fig:pretrain}, 
%
we first train a single main sparse autoencoder $\mathcal{A}_{\text{pre}}$ as shown in the upper part of Fig.~\ref{fig:pretrain}, for the entire dataset using the loss function described at equation \ref{total_loss}. 
After training the main sparse autoencoder $\mathcal{A}_{\text{pre}}$, one off-the-shelf cluster algorithm such as AHC or k-Means, is utilized to obtain initial pseudo-labels $\boldsymbol{{P}}^{[0]}$ from the learned latent representation of the sparse autoencoder $\mathcal{A}_{\text{pre}}$.
Next, we initialize the parameters of k-sparse autoencoders by sequentially training the $j$-th sparse autoencoder $\mathcal{A}_j$ with the subset of points such that $\boldsymbol{\boldsymbol{P}}^{[0]}[c=j]$, as shown in the lower part of Fig.~\ref{fig:pretrain}, where $c$ denotes the cluster index, $j=1, 2, ..., k$. Notably, the training process of k-sparse autoencoders also use Equation $\ref{total_loss}$ as the loss function. 

The next Main-training step is described in Fig. \ref{fig:mae}. This step involves the joint optimization of the k-sparse autoencoders with initialized parameters obtained from the Pre-training step, and the predicted probabilities from the gating projection. 
Given k-sparse autoencoders $\{\mathcal{A}_1(\theta_1),\mathcal{A}_2(\theta_2), ..., \mathcal{A}_k(\theta_k)\}$, where $\theta_j$ is the parameters of encoder ($\mathcal{E}_j$) and decoder ($\mathcal{D}_j$) of sparse autoencoder $\mathcal{A}_j$, $j=1,2,...,k$, and the parameters ($\boldsymbol{W}$, $\boldsymbol{B}$) of the gating projection $\mathcal{G}$, the main objective function of the proposed Mix-SAE network for one batch of $N$ samples [$\boldsymbol{x}^{(1)}, \boldsymbol{x}^{(2)}, ..., \boldsymbol{x}^{(N)}$] is defined as:
\begin{equation}
    \mathcal{L}_{\text{main}}(\theta_1, \theta_2, ..., \theta_k, \boldsymbol{W}, \boldsymbol{B}) = \mathcal{L}_{\text{rec}} + \alpha \mathcal{L}_{\text{ent}}
    \label{main_obj}
\end{equation}
where $\alpha$ is the parameter to constrain the effect of both terms on the main objective function.

The term $ \mathcal{L}_{\text{rec}}$ is the weighted sum of reconstruction error over k-sparse autoencoders. 
This term ensures that the sparse autoencoders could have information on inter-cluster reconstruction error to further strengthen feature learning within their own clusters. We define this term as:
\begin{equation}
\begin{aligned}
& \mathcal{L}_{\text{rec}} = -\frac{1}{N} \sum^{N}_{i=1} \sum^{k}_{j=1} \hat{p}^{(i)}_j \exp{\left[-\frac{1}{2}\left(\boldsymbol{x}^{(i)} - \mathcal{D}_j(\mathcal{E}_j(\boldsymbol{x}^{(i)}))\right)^2\right]} \\
& ~~~~~~~~~~~~~~\textrm{s.t.} ~~\sum^{k}_{j=1} \hat{p}^{(i)}_j = 1, \quad \forall i = 1, 2, \ldots, N.
\end{aligned}
\end{equation}
where $\mathcal{D}_j(\mathcal{E}_j(\boldsymbol{x}^{(i)}))$ is the output of the $j$-th sparse autoencoder given the input sample $\boldsymbol{x}^{(i)}$; the probability $\hat{p}^{(i)}_j$, which is computed from ($\boldsymbol{W}$, $\boldsymbol{B}$) in Equation~\ref{gate}, is the weight from the gating projection assigned to the $j$-th reconstruction loss.

The term  $\mathcal{L}_{\text{ent}}$ is referred to as the pseudo-label guided supervision loss. We denote the pseudo-labels for one batch of $N$ samples at epoch $t$ as: $\mathcal{\mathbf{P}}^{[t]} = [\boldsymbol{p}^{[t]}_1, \boldsymbol{p}^{[t]}_2,..., \boldsymbol{p}^{[t]}_N]$, where $\boldsymbol{p}^{[t]}_i \in \mathbb{R}^{k}$. The supervision loss is defined as the Cross-Entropy loss between the pseudo-labels $\boldsymbol{p}^{[t_{\text{u}}]}$ previously updated at epoch $t_{\text{u}}$ and the prediction of the gating projection $\hat{\boldsymbol{p}}^{[t_{\text{u}}]}$ at the current epoch $t$:

\begin{equation}
    \mathcal{L}_{\text{ent}} = -\frac{1}{N}\sum^{N}_{i=1}\boldsymbol{p}^{[t_{\text{u}}]}_i\log{\hat{\boldsymbol{p}}^{[t]}_i}
    \label{ent}
\end{equation}

\begin{table}[t]
    \centering
    \caption{Mix-SAE Deep Clustering Network}
    \vspace{-0.3cm}
    \label{tab:algo}
    \scalebox{0.765}{
    \begin{tabular}{l} \\ \hline
         \textbf{Algorithm 1}: Mix-SAE mini-batch training strategy \\ \hline
         \textbf{\underline{Input:}}  One batch of $N$ points $\boldsymbol{X}$=$\{\boldsymbol{x}^{(i)}\}^{N}_{i=1} \in \mathbb{R}^m$. \\
          \textbf{\underline{Output:}} One of $k$ cluster labels for $N$ input points.\\ 
         \textbf{\underline{Components}}: \\ - A set of k-autoencoders: $\{\mathcal{A}_1, \mathcal{A}_2, ..., \mathcal{A}_k\}$ 
          \vspace{0.1cm}\\
          \,\,\,\,\,\,\,\,\,\,\,\,\,\,\,\,
        $\boldsymbol{x}\rightarrow\bar{\boldsymbol{x}}_{j}=\mathcal{D}_j(\mathcal{E}_j(\boldsymbol{x})), \,\,\,\,\, j = 1, 2, ..., k.$
         \vspace{0.1cm}\\
         - The Gating Projection $\mathcal{G} $ that produces pseudo-labels and  assigns input to suitable autoencoders.
         \vspace{0.1cm}\\
          \,\,\,\,\,\,\,\,\,\,\,\,\,\,\,\,
           $\boldsymbol{p} = Softmax(\boldsymbol{W}\boldsymbol{x} + \boldsymbol{b}) \in \mathbb{R}^c$.
         \vspace{0.1cm}\\
        
         \textbf{• \underline{Pre-training:}}\\
         -  Train a single autoencoder $\mathcal{A}_{\text{pre}}$ for the entire dataset with the objective function (\ref{total_loss}).\\
         - Use one off-the-shelf cluster algorithm to initialize pseudo-labels $\boldsymbol{P}^{[0]}$ for the entire dataset.\\
         \textbf{for} $j=1$ to $k$ \textbf{do}:\\
         \,\,\,\,\, Train $j$-th sparse autoencoder with data points $\boldsymbol{P}^{[0]}[c=j]$.\\
         \textbf{end for}\\ 

         \textbf{• \underline{Main-training:}}\\
         \textbf{for} $t=1$ to $T$ \textbf{do}:\\
         \,\,\,\,\, Train the set of k sparse auencoders and the gating projection $\mathcal{G}$ jointly using the main objective \\
         \,\,\,\,\,  function (\ref{main_obj}).\\
         \,\,\,\,\, \textbf{if} $t$ mod $\tau = 0$ \textbf{then}:\\
         \,\,\,\,\,\,\,\,\,\,\,\,\,\,\, Update new pseudo-labels $\boldsymbol{P}^{[t_{\text{u}}]}$ for the batch $\boldsymbol{X}$:
         \vspace{0.1cm}\\
          \,\,\,\,\,\,\,\,\,\,\,\,\,\,\,\, \,\,\,\,\,\,\,\,\,\,\,\,\,\,\,\,
         $t_{\text{u}}\leftarrow t$ \vspace{0.1cm}\\
          \,\,\,\,\,\,\,\,\,\,\,\,\,\,\,\, \,\,\,\,\,\,\,\,\,\,\,\,\,\,\,\,
         $\boldsymbol{P}^{[t_{\text{u}}]} = \underset{\text{axis = 1}}{\argmax} \left[Softmax(\boldsymbol{W}\boldsymbol{X} +\mathbf{B})\right]$ \\
        
        
        \\ 
         \textbf{\underline{Get final cluster result:}} Get the final cluster result for the batch $\boldsymbol{X}$ via the gating projection $\mathcal{G}$:\\
         \vspace{0.1cm}\\
          \,\,\,\,\,\,\,\,\,\,\,\,\,\,\,\, \,\,\,\,\,\,\,\,
         
        $\hat{\boldsymbol{P}} = \underset{\text{axis = 1}}{\argmax} \left[Softmax\left(\boldsymbol{W}\boldsymbol{X}+\boldsymbol{B}\right)\right]$\\ \\
        \hline
    \end{tabular}
    }
    \vspace{-0.4cm}    
\end{table}
The entropy loss $\mathcal{L}_{\text{ent}}$ uses pseudo-labels to provide additional learning signals, simulating a semi-supervised setting. This guides the model towards correct clustering and enhances feature learning~\cite{min2023pseudo}. Notably, pseudo-labels are periodically updated after $\tau$  epochs during optimization by the predictions of the gating projection $\mathcal{G}$ at the current epoch $t$ using Equation~\ref{gate}. This process aims to reinforce reliable pseudo-labels while correcting noisy ones over time.

After the Main-training step, final cluster label can be inferred via the gating projection $\mathcal{G}$. Given each data sample $\boldsymbol{x}$, the probability vector $\hat{\boldsymbol{p}}$ is calculated using equation \ref{gate}. Then, the cluster label is determined as: 
\begin{equation}
    \hat{c} = \argmax \,\hat{\boldsymbol{p}} = \underset{j=1,2, .. k}{\argmax}\, \hat{p}(c = j|\boldsymbol{x})
    \label{get_label}
\end{equation}
Overall, steps in the training strategy of our proposed Mix-SAE clustering network can be summarized in Table \ref{tab:algo}.
\renewcommand{\arraystretch}{1.6}

\section{Experimental Settings And Results}
\subsection{Evaluation Datasets}
To evaluate the performance and generalization of our proposed system to diverse data sources, we gather data from two benchmark corpora CALLHOME~\cite{callhome_eng},\cite{callhome_german},\cite{callhome_spanish} and CALLFRIEND~\cite{callfriend_french}, \cite{callfriend_spanish}. Each corpus includes various language subsets like English, German, French, Spanish, and Japanese, with multiple telephone conversations from different sources. For evaluation, we use two-speaker subsets of the above benchmark corpora (the most common case in telephone call applications), to form a combined dataset called SD-EVAL.
The SD-EVAL dataset comprises 127 recordings totaling around 6.35 hours and is divided into four language-specific subsets: English (EN), Spanish (SPA), German (GER), and French (FR). Each subset has 25 to 35 recordings, each lasting 2 to 5 minutes.

\subsection{Evaluation Metrics}
We evaluated the proposed sysetm with diarization error rate
(DER). 

\subsection{Experimental settings}
The proposed method was implemented with deep
learning framework PyTorch \cite{Pytorch}. The network architecture consists of autoencoders with hidden layers [256, 128, 64, 32] for the encoder and mirrored for the decoder, using Leaky ReLU activation and Batch Normalization followed each hidden layer. The latent vector size is also $k$ (equal to the number of speakers), with mini-batch size $N=16$. We use k-Means$^{++}$~\cite{kmeans++} to initialize pseudo-labels in the Pre-training step.

Regarding hyperparameters, we set sparsity parameter $\rho = 0.2$, sparsity constraint $\beta=0.01$, pseudo-label supervision $\alpha=1$. The training process uses learning rate $0.001$ and weight decay $5\times10^{-4}$. The Pre-training step involves 50 epochs for the main autoencoder $\mathcal{A}_{\text{pre}}$ and 20 epochs for each of k-sparse autoencoders. The Main training step runs for 29 epochs and updates pseudo-labels after 10 epochs.

\begin{table*}[t]
\centering
\caption{Diarization Error Rate (DER) (\%) of different systems on SD-EVAL dataset (Whisper version: Tiny, no tolerance)}
\label{system_performance}
\scalebox{0.57}{
\begin{tabular}{l cccc cccc cccc cccc cccc}
\toprule
 & \multicolumn{4}{c}{$W=0.2$ s} &
 \multicolumn{4}{c}{$W=0.4$ s} &
 \multicolumn{4}{c}{$W=0.6$ s} &
 \multicolumn{4}{c}{$W=0.8$ s} &
 \multicolumn{4}{c}{$W=1.0$ s} \\
\cmidrule(lr){2-5} \cmidrule(lr){6-9} \cmidrule(lr){10-13} \cmidrule(lr){14-17} \cmidrule(lr){18-21}
\textbf{Methods}     & EN   & FR  & GER & SPA   
& EN   & FR  & GER & SPA   
& EN   & FR  & GER & SPA 
& EN   & FR  & GER & SPA  
& EN   & FR  & GER & SPA   \\
\midrule[\heavyrulewidth]

k-Means        & 44.77 & 51.42 & 49.11  & 48.25  
              & 43.75 & 51.92 &43.84 & 47.08
              & 38.72 & 46.88 & 40.97 & 42.77
              & 40.23  & 46.61 & 44.11 & 44.38
              & 42.06  & 47.72 & 46.13 & 44.66
              \\
AHC           & 38.42 & 46.72 & 41.41  & 42.93  
              & 47.64  & 52.81 & 46.33 & 50.69 
              & 40.50  & 48.69 & 42.90& 43.15
              & 38.55  & 45.91 & 43.02 & 43.44
              & 42.91  & 47.81 & 47.63 & 44.80
              \\
SpectralNet \cite{spectralnet}                         & 36.18 & 44.62 &              40.02  & 46.03
        & 40.44  & 51.63 & 41.22 & 47.52
         & 37.06  & 44.68 & 41.29 & 42.69
        & 36.11  & 44.67 & 44.16 & 46.42
        & 41.88  & 46.08 & 44.31 & 47.23
        \\
DCN \cite{dcn}       
              & 32.15 & \textbf{35.77} & 36.51  & 36.98  
              & 37.42  & 38.92& 42.17 & 43.01
              & 32.08  & 37.57& 38.84 & 40.77
              & 33.02  & 43.72 & 44.23 & 40.55 
              & 40.17  & 45.96 & 40.21 & 38.51
              \\
DAMIC \cite{damic}      
             & 27.78 & 36.22 & 36.93 & 35.21  
              & 27.97  & 35.96 & 36.14 & 35.11
              & 28.11 & 36.67 & 34.66& 33.31
              & 27.22    & \textbf{36.91} & 34.78 & 34.22 
              & 26.95  & \textbf{36.91} & 36.11 & 34.65
              \\
k-DAE \cite{kdae}       
              & 29.12& 37.91 & 41.23 & 37.00 
              & 30.53  & 39.81 & 37.10 & 37.29
              & 32.72  &38.84 & 34.96& 35.23
              & 33.33 & 38.55 & 34.24 & 35.51 
              & 30.36  & 37.32 & 36.22 & 35.02
              \\
\hline
\textbf{Mix-SAE-V1}      
              & 32.18 & 38.61 & 36.07  & 36.78  
              & 29.02  & \textbf{35.92} & 36.51 & 35.04 
              & 27.28 & 37.01 & 34.98 & 34.03 
              & 27.90  & 37.51 & 34.42 & 33.83 
              & 28.00  & 37.88 & 36.18 & 34.29
              \\
\textbf{Mix-SAE-V2}        
              & 28.72 & 43.22 & 40.66& 36.32 
              & 29.62  & 40.07 & 36.71 & 35.72 
              & 27.81  & 36.83 & 34.90 & 33.54 
              & 27.98 & 39.68 & 34.62& 33.21
              & 27.93  & 38.05 & 36.73 & \textbf{33.82} 
              \\
\bottomrule
\textbf{Mix-SAE}        
              & \textbf{26.51} & 36.12 & \textbf{35.00} & \textbf{34.91}
              &  \textbf{26.88}  &37.30 &  \textbf{35.64} & \textbf{34.33} 
              &  \textbf{27.08}  & 36.70 &  \textbf{34.55} & \textbf{32.82}
              & \textbf{27.24}  & 38.39 & \textbf{34.17} & \textbf{32.03} 
              & \textbf{26.85}  & 37.57& \textbf{35.33} & \textbf{33.82} 
              \\
\bottomrule
\end{tabular}
}
\end{table*}

\subsection{Results and Discussion}
\textbf{Speaker clustering methods:}
We evaluate several speaker clustering methods using embeddings from the tiny Whisper model, including k-Means, Agglomerative Hierarchical Clustering (AHC), SpectralNet, autoencoder-based methods such as DCN, DAMIC, k-DAE, and our proposed Mix-SAE. Experiments were conducted with segment sizes ($W$) ranging from 0.2s to 1.0s. As shown in Table \ref{system_performance}, Mix-SAE consistently outperforms other methods, achieving the best performance in English with a DER of 26.51\%. This can be attributed to high-quality embeddings from Whisper's extensive English training data. While other methods, especially autoencoder-based ones like DCN and k-DAE, show variability with segment size, Mix-SAE remains stable across different $W$ values, demonstrating its efficiency in capturing speaker features from variable-length segments (e.g. the proposed system achieves DER scores of 26.51\%, 26.88\%, 27.08\%, 27.24\%, 26.85\% on English and 35.00\%, 35.64\%, 35.55\%, 34.17\%, 34.55\% on German with $W$ = 0.2, 0.4, 0.6, 0.8, 1.0, respectively).
\begin{figure}[t]
    \centering
  \subfloat[]{%
       \includegraphics[width=0.46\linewidth]{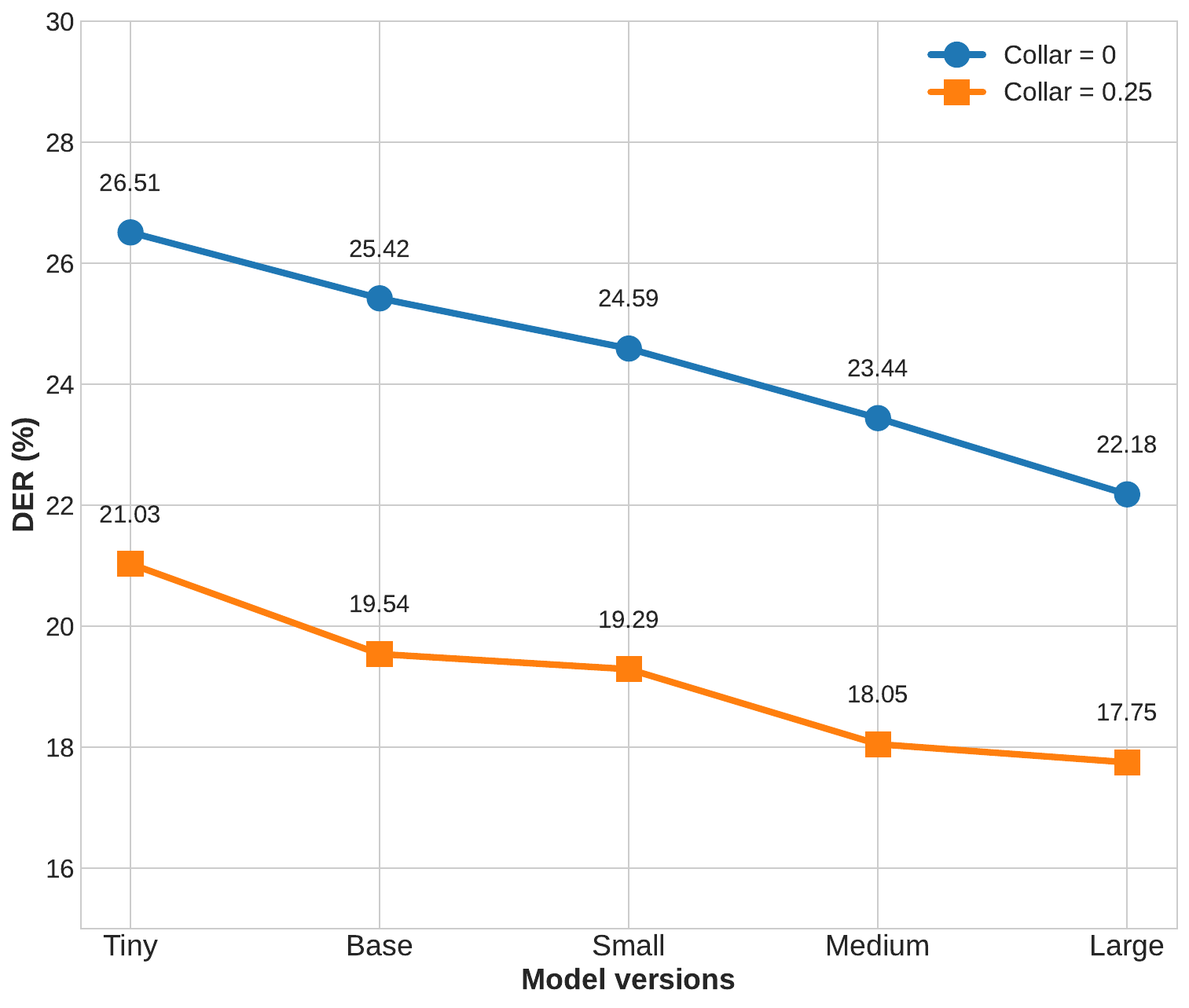}}
    \hspace{0.1cm}
  \subfloat[]{%
        \includegraphics[width=0.5\linewidth]{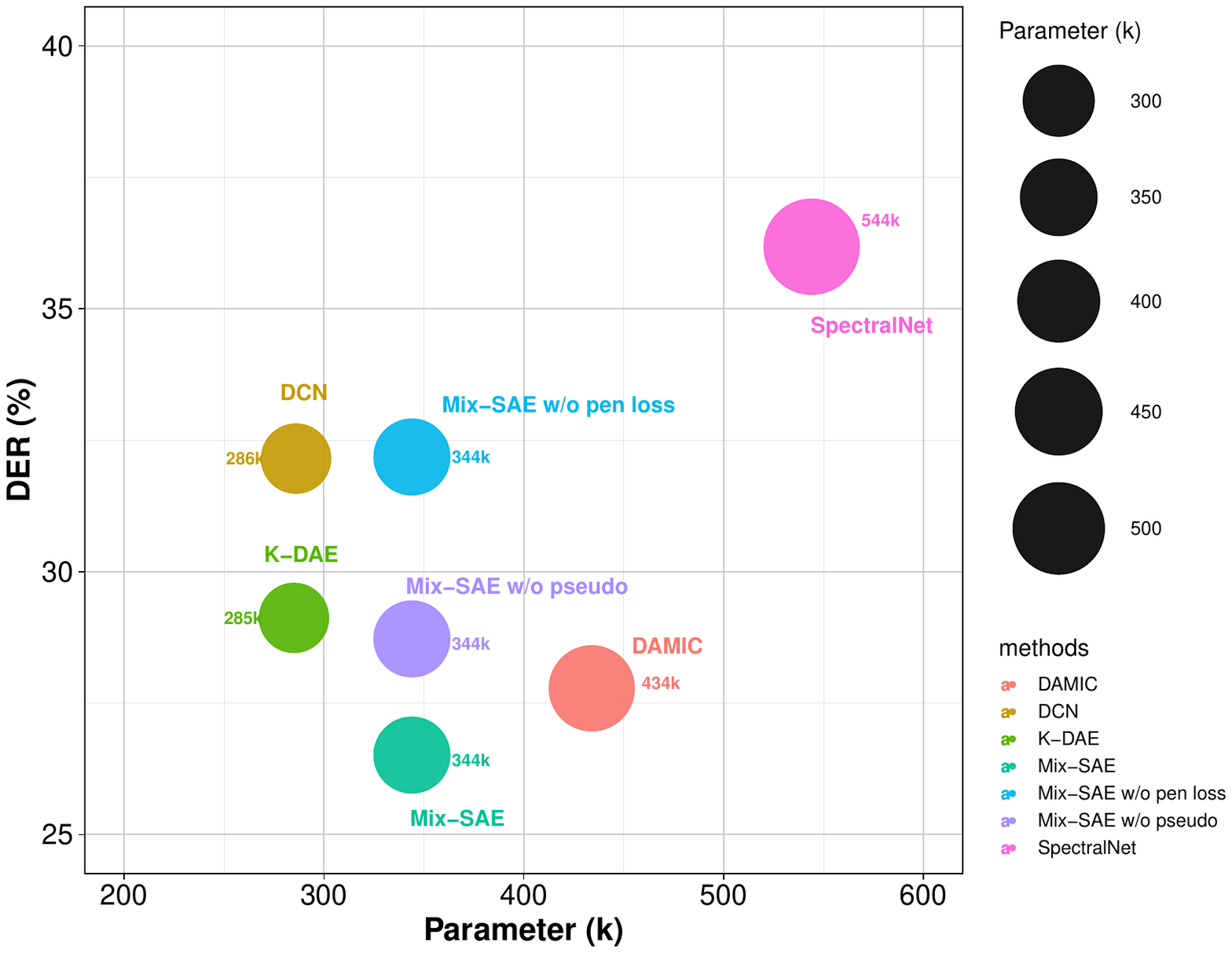}}
  \caption{\textit{Evaluation: (a) DER scores using speaker embeddings from different Whisper versions; (b) Compare DER score versus complexity across deep clustering methods}}
 \label{speaker_ab}
\end{figure}
For an ablation study, we establish two other systems: Mix-SAE-V1 (Mix-SAE without sparsity loss in equation \ref{total_loss}), Mix-SAE-V2 (Mix-SAE w/o pseudo-label loss in equation \ref{main_obj}). 
Results in Table \ref{system_performance} demonstrate the role of both sparsity loss and pseudo-label loss in improving the overall performance. 
For instance, an improvement of 5.67\% and 2.21\% is obtained in the case of English with $W$ = 0.2s when Mix-SAE is compared to Mix-SAE-V1 and Mix-SAE-V2, respectively. 
\begin{figure}[h]
    \centering
  \subfloat[English ($W=0.2s$)]{%
       \includegraphics[width=0.35\linewidth]{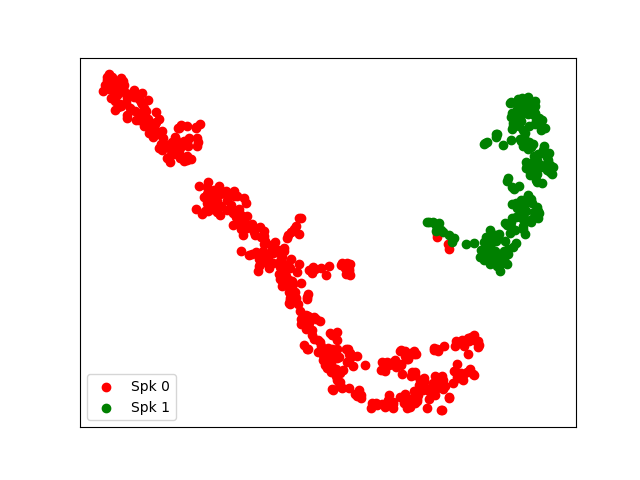}}
    \hspace{0.3cm}
  \subfloat[French ($W=0.2s$)]{%
        \includegraphics[width=0.35\linewidth]{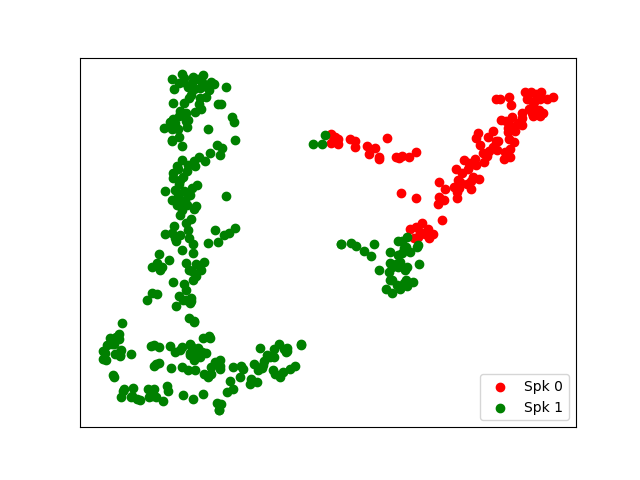}}
    \\
  \subfloat[German ($W=0.2s$)]{%
        \includegraphics[width=0.35\linewidth]{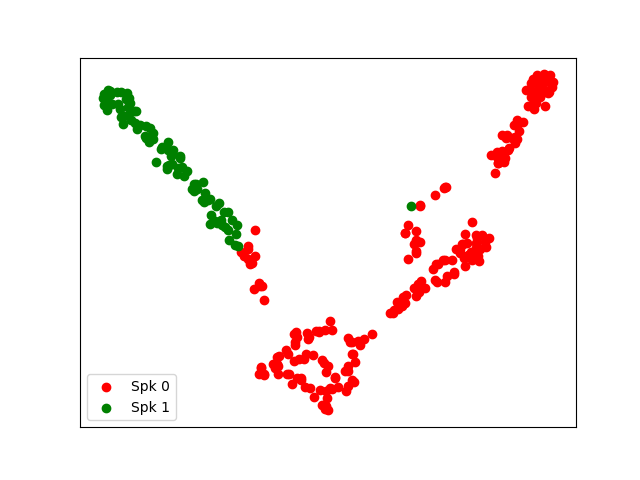}}
    \hspace{0.3cm}
  \subfloat[Spanish ($W=0.2s$)]{%
        \includegraphics[width=0.35\linewidth]{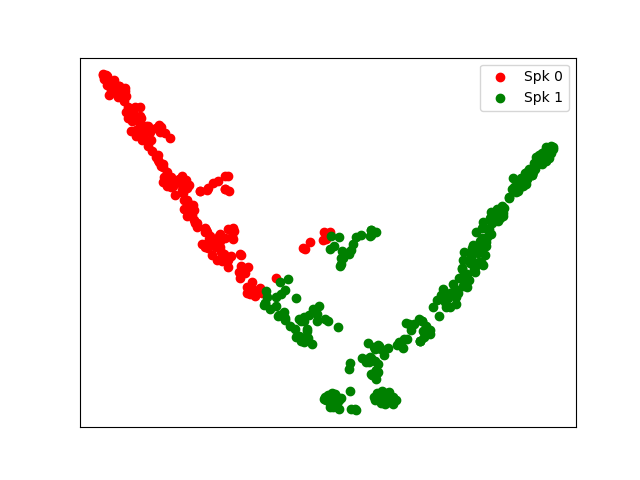}}
  \caption{\textit{t-SNE visualization of speaker embeddings after the pre-training step (Whisper version: 
 Tiny)}}
 \label{speaker_visualization}
\end{figure}

\textbf{The quality of speaker embeddings:} 
We assessed the impact of speaker embeddings on diarization performance, as shown in Fig. \ref{speaker_ab}a, using different versions of the Whisper model (Tiny, Base, Small, Medium, Large) with $W$ set to 0.2s in English. Larger Whisper models provided superior embeddings, leading to better performance, with the best DER score of 17.75\% (0.25s tolerance). This
highlights the potential of using general-purpose like Whisper for multilingual and unsupervised speaker diarization systems as well as integrating speaker diarization as a component in Whisper-based speech analysis applications.


\textbf{The model complexity:} 
Fig. \ref{speaker_ab}b shows the trade-off between model complexity and diarization performance (DER) across deep clustering methods. Our Mix-SAE achieves 26.51\% DER with 334k parameters, striking a good balance between accuracy and efficiency. Additionally, when combined with Whisper Tiny (39M), the system is promising for integration into edge devices for sound applications~\cite{wp_embed1}, \cite{wp_embed2}.

\textbf{Visualization and the effect of Pre-training step:} We visualized 2-speaker embeddings after the Pre-training step in our Mix-SAE by applying t-SNE. 
As Fig.~\ref{speaker_visualization} shows, the sparse autoencoders effectively learn underlying patterns from extracted speaker embeddings and map them into latent space where the embeddings of two speakers were relatively well-separated.
These clustering results serve as pseudo-labels for optimizing the deep clustering network at the next Main-training step. 


\section{Conclusion}
This paper has presented an unsupervised speaker diarization system for multilingual telephone call applications. 
In this proposed system, the traditional feature extractor was replaced with the Whisper encoder, benefiting from its robustness and generalization on diverse data. 
Additionally, the Mix-SAE network architecture was also proposed for speaker clustering. Experimental results demonstrate that our Mix-SAE network outperforms other compared clustering methods. The overall performance of our system highlights the effectiveness of our approach when exploring Whisper embedding for the  diarization task to develop unsupervised speaker diarization system in the contexts of limited annotated training data. Furthermore, the results also enhances the system's ability to integrate into Whisper-based multi-task speech analysis application. Overall, this work indicates a promising direction toward developing generalized speaker diarization systems based on general-purpose models in future work.
\section*{Acknowledgment}
\begin{wrapfigure}{r}{2.5cm}
\centering
\vspace{-0.3cm}
\label{wrap-fig:1}
\vspace{-0.5cm}
\includegraphics[width=2.cm, height = 1.8cm]{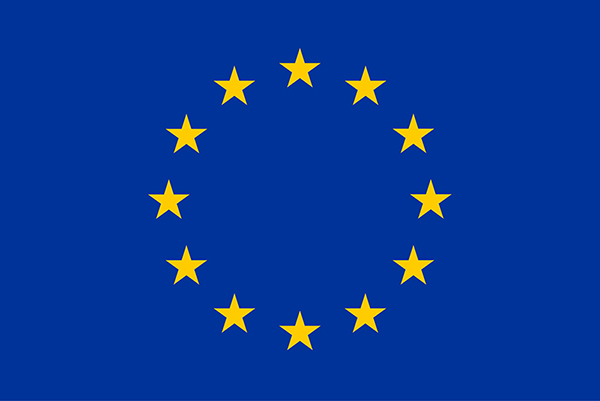}
\end{wrapfigure} 

The work described in this paper is performed in the H2020 project STARLIGHT (“Sustainable Autonomy and Resilience for LEAs using AI against High Priority Threats”). This project has received funding from the European Union’s Horizon 2020 research and innovation program under grant agreement No 101021797.

%
%
%





\end{document}